\documentclass[letterpaper,twoside,10pt]{article}
\usepackage{amssymb}
\usepackage{amsmath}
\usepackage{latexsym}
\usepackage{fancyheadings}
\newlength{\enviropost}
\newcommand{\be}{\begin{equation}}
\newcommand{\ee}{\end{equation}}
\newcommand{\ble}[1]{\begin{equation} \label{#1}}
\newcommand{\bae}{\begin{eqnarray}}
\newcommand{\eae}{\end{eqnarray}}
\newcommand{\fle}[2]%
{\vspace{1.5ex}
\be
\label{#1}
\mbox{%
\setlength{\fboxsep}{3ex}%
\framebox{$\dss #2 $}}
\ee} 
\newcommand{\flec}[2]%
{\vspace{1.5ex}
\be
\label{#1}
\mbox{%
\setlength{\fboxsep}{3ex}%
\framebox{$\dss #2 $}}
\, \, \,  ,
\ee} 
\newcommand{\flep}[2]%
{\vspace{1.5ex}
\be
\label{#1}
\mbox{%
\setlength{\fboxsep}{3ex}%
\framebox{$\dss #2 $}}
\, \, \, .
\ee} 
\newcommand{\nn}{\nonumber}
\newcommand{\ff}{\nn \\}
\newcommand{\fe}{& = &}

\newtheorem{state}{S$\! \!$}
\newtheorem{defin}{D$\! \!$}
\newtheorem{exatitle}{Example}
\newtheorem{problemdef}{Problem}
\newtheorem{soldef}{Solution}
{\noindent \textsc{Proof}:\ }% 
{\hfill $\blacksquare$  \vspace{\enviropost} \\}
{\begin{soldef} 
\label{#2:Sol} 
#1
\marginpar[(\textbf{\pageref{#2:Pro}})]{(\textbf{\pageref{#2:Pro}})}
\end{soldef}}% 
{\hfill $\Box$ \vspace{.5\enviropost} \\}
{\begin{exatitle} \label{#2} #1 \end{exatitle}}% 
{\hfill $\Box$ \vspace{\enviropost} \\}
{\begin{exatitle} \label{#2} #1 \end{exatitle}}%
{\hfill $\Box$ \\}
{\begin{problemdef} 
\label{#2:Pro} 
#1 
\marginpar[(\textbf{\pageref{#2:Sol}})]{(\textbf{\pageref{#2:Sol}})}
\end{problemdef}}% 
{\hfill  \vspace{.5\enviropost} \\}
{\noindent {\bf Aside:} \small}% 
{\hfill \rule[-3mm]{0mm}{0mm}$\Diamond$\\}
{%
\vspace{3mm}  
\begin{center}
\begin{minipage}{.8\textwidth} 
\begin{state} 
\label{#1}%
}% 
{%
\end{state}
\end{minipage}
\end{center}
\vspace{3mm}
}
{%
\vspace{3mm}  
\begin{center}
\begin{minipage}{.8\textwidth} 
\begin{defin} 
\label{#1}%
}% 
{%
\end{defin}
\end{minipage}
\end{center}
\vspace{3mm}
}
\newcommand{\eg}{\hbox{\em e.g.{}}}

\newcommand{\ie}{\hbox{\em i.e.{}}}

\newcommand{\rhs}{\hbox{r.h.s.{}}}

\newcommand{\calG}{\mathcal{G}}

\newcommand{\calL}{\mathcal{L}}

\newcommand{\papertitle}{%
Linear Form of
3-scale Special Relativity Algebra\\[1.5mm]
and the Relevance of Stability%
}
\newcommand{\runningtitle}{%
Linear Form of
3-scale Special Relativity Algebra \ldots%
}
\newcommand{\paperauthor}{%
C.{} Chryssomalakos and E.{} Okon%
}
\begin{document}
%\initfloatingfigs
%%%%%%%%%%%%%%%%%%%%%%%%%%%%%%%%%%%%%%%%%%%%%%%%%%%%%%%%%%%%%%
%%%%%%%%%%%%%%%%%%%%%%%%%%%%%%%%%%%%%%%%%%%%%%%%%%%%%%%%%%%%%%
%%%%%%%%%%%%%%%%%%%%%%%%%%%%%%%%%%%%%%%%%%%%%%%%%%%%%%%%%%%%%%
% Titlepage
%%%%%%%%%%%%%%%%%%%%%%%%%%%%%%%%%%%%%%%%%%%%%%%%%%%%%%%%%%%%%%
\begin{titlepage}
\vspace*{-1cm}
\begin{flushright}
\textsf{}
%\\
%\textsf{ICN-UNAM-yy/pp}
\\
\mbox{}
\\
\textsf{\today}
\\[3cm]
\end{flushright}
%%%%%%%%%%%%%%%%%%%%%%%%%%%%%%%%%%%%%%%%%%%%%%%%%%%%%%%%%%%%%%
%%%%%%%%%%%%%%%%%%%%%%%%%%%%%%%%%%%%%%%%%%%%%%%%%%%%%%%%%%%%%%
%%% TITLE, AUTHORS
%%%%%%%%%%%%%%%%%%%%%%%%%%%%%%%%%%%%%%%%%%%%%%%%%%%%%%%%%%%%%%
%%%%%%%%%%%%%%%%%%%%%%%%%%%%%%%%%%%%%%%%%%%%%%%%%%%%%%%%%%%%%%
%\begin{center}
%%%%%%%%%%%%%%%%%%%%%%%%%%%%%%%%%%%%%%%%%%%%%%%%%%%%%%%%%%%%%%
\renewcommand{\thefootnote}{\fnsymbol{footnote}}
\begin{LARGE}
\bfseries{\sffamily \papertitle}
\end{LARGE}

\noindent \rule{\textwidth}{.6mm}

\vspace*{1.6cm}

\noindent \begin{large}%
\textsf{\bfseries%
\paperauthor
}
\end{large}

%\vspace*{.1cm}

\phantom{XX}
\begin{minipage}{.8\textwidth}
\begin{it}
\noindent Instituto de Ciencias Nucleares \\
Universidad Nacional Aut\'onoma de M\'exico\\
Apdo. Postal 70-543, 04510 M\'exico, D.F., M\'EXICO \\
\end{it}
\texttt{chryss@nuclecu.unam.mx, eliokon@nuclecu.unam.mx
\phantom{X}}
\end{minipage}
\\

\vspace*{3cm}
%%%%%%%%%%%%%%%%%%%%%%%%%%%%%%%%%%%%%%%%%%%%%%%%%%%%%%%%%%%%%%
%%% ABSTRACT
%%%%%%%%%%%%%%%%%%%%%%%%%%%%%%%%%%%%%%%%%%%%%%%%%%%%%%%%%%%%%%
\noindent
\textsc{\large Abstract: }
We show that the algebra of the recently proposed Triply 
Special Relativity can be brought to a linear (\ie, Lie) form by a
correct identification of its generators. The resulting Lie algebra is
the
stable form proposed by Vilela Mendes a decade ago, itself a
reapparition of Yang's algebra, dating from 1947. As a
corollary we assure that, within the Lie algebra framework, there is no 
Quadruply Special Relativity. 
\end{titlepage}
\setcounter{footnote}{1}
\renewcommand{\thefootnote}{\arabic{footnote}}
\setcounter{page}{2}
%%%%%%%%%%%%%%%%%%%%%%%%%%%%%%%%%%%%%%%%%%%%%%%%%%%%%%%%%%%%%%%%%
%%%%%%%%%%%%%%%%%%%%%%%%%%%%%%%%%%%%%%%%%%%%%%%%%%%%%%%%%%%%%%%%%
%%%%%%%%%%%%%%%%%%%%%%%%%%%%%%%%%%%%%%%%%%%%%%%%%%%%%%%%%%%%%%%%%
%\noindent \rule{\textwidth}{.5mm}
%
%\tableofcontents
%
%\noindent \rule{\textwidth}{.5mm}
%%%%%%%%%%%%%%%%%%%%%%%%%%%%%%%%%%%%%%%%%%%%%%%%%%%%%%%%%%%%%%%%
%%%%%%%%%%%%%%%%%%%%%%%%%%%%%%%%%%%%%%%%%%%%%%%%%%%%%%%%%%%%%%%%%
\section{Introduction}
\label{Intro}
%%%%%%%%%%%%%%%%%%%%%%%%%%%%%%%%%%%%%%%%%%%%%%%%%%%%%%%%%%%%%%%%
%%%%%%%%%%%%%%%%%%%%%%%%%%%%%%%%%%%%%%%%%%%%%%%%%%%%%%%%%%%%%%%%%
During the last few years, a significant number of papers has
focused on the question of 
deforming the Lie algebra of standard quantum relativistic kinematics. 
By the latter we refer to the Poincar\'e algebra, extended by
the promotion of the coordinates $X^\mu$ to generator status. The main
motivation behind this effort is the inference of an
algebraic signature of quantum gravity. Accordingly, the
first step taken, is the postulation of the Heisenberg
commutation relations among $X^\mu$ and the momenta $P_\mu$,
while the four-vector nature of the former is retained.
The new invariant scale
introduced as a result of the deformation 
(in addition to those given by the velocity of
light and Planck's constant), is generally identified with the 
Planck scale.
The resulting spacetime is non-commutative, the characteristic
length of the non-commutativity given by the Planck length.
Quite recently, an additional deformation was
proposed~\cite{Kow.Smo:04}, that,
roughly speaking, does the same to the momentum sector. The new
(fourth) invariant scale introduced is related to the
cosmological constant, and shows up in the non-commutativity of
the momenta. 

In the initially  proposed deformations, 
the commutators of the generators were expressed as 
general analytic functions of the same, rather than linear ones,
\ie, the Lie algebra framework
was abandoned. It was later realized that, by suitable
(nonlinear) redefinitions of the generators, linearity could be
restored, \ie,
the proposed deformations lived, after all, within the 
(suitably completed) universal enveloping
algebra of a Lie algebra (albeit, in some cases, with a different 
coproduct). 
In the recent additional deformation mentioned above, the
imposition of the Jacobi identities results, according to the
authors of~\cite{Kow.Smo:04}, to non-linearity of the
$x-p$ commutation relations. 
Since the rest of the
commutation relations are linear, it is not at all clear that this
non-linearity can be also made to disappear by a 
redefinition of the generators.
This would seem to suggest that the introduction of this
new invariant scale in Special Relativity can only be achieved at
the cost of parting with the familiar, and powerful,
Lie algebra machinery.  
Our first aim in this note is to
point out that this is not the case. A proper identification 
of the generators permits to close
the algebra linearly. The point, we think, is particularly important. 
It shows that {\em invariant scales can be consistently introduced
in Special Relativity within the Lie algebra framework}.  
Our wider, and, by far, most important, 
aim though is to bring to the attention of the authors 
in the field the relevance of the formal Lie algebra
deformation theory and the related concept of Lie algebra stability. 
With this in mind, we comment on the similarities between the 
non-linear algebras that have been proposed and the stable Lie algebra
put forth by Vilela Mendes in~\cite{Vil:94} more than a decade ago%
\footnote{%
This algebra has actually first appeared in the work of
Yang~\cite{Yan:47} --- see Sect.{}~\ref{RwOA}%
}.  

Before delving into these matters in detail, we wish to
make two comments. The first one concerns the scope of the 
present work. As a
rule, the above mentioned deformations are Hopf algebra
deformations.
 In this short
note, we deal exclusively with their algebraic sector --- a 
thorough analysis of their full Hopf
algebra structure, as well as a detailed presentation of the
Lie algebra stability point of view and the physical
interpretation of the resulting stable algebra 
will be the subject of a longer article,
presently in preparation. 

The second comment concerns nomenclature. 
The generally used term for
this field is ``Doubly Special Relativity'' (DSR) --- recently
promoted to ``Triply''. As explained in sufficient detail
in~\cite{Ahl:03}, this is a misnomer. The ``Special'' of
``Special Relativity'' refers to the restriction on the coordinate
transformations considered, not to the existence of an invariant
scale, therefore, the multiple-scale relativities proposed are
as (singly) Special as the original one. 
We think the above term is conceptually inappropriate
enough to warrant its abolishment, and propose as an
alternative the ``$n$-scale Special Relativity'' ($n$-SSR) of
the title. As explained below, $n$=0,1,2,3 are the only
possibilities within the Lie algebra framework. We emphasize
at this point that $n>1$ does not imply non-linearity ---
non-linear deformations will be referred to explicitly, \eg,
the algebra
of~\cite{Kow.Smo:04} is a non-linear 3-SSR.
%%%%%%%%%%%%%%%%%%%%%%%%%%%%%%%%%%%%%%%%%%%%%%%%%%%%%%%%%%%%%%%%
%%%%%%%%%%%%%%%%%%%%%%%%%%%%%%%%%%%%%%%%%%%%%%%%%%%%%%%%%%%%%%%%%
\section{The Linear Form of ``Triply Special Relativity''}
\label{TLFTSR}
%%%%%%%%%%%%%%%%%%%%%%%%%%%%%%%%%%%%%%%%%%%%%%%%%%%%%%%%%%%%%%%%
%%%%%%%%%%%%%%%%%%%%%%%%%%%%%%%%%%%%%%%%%%%%%%%%%%%%%%%%%%%%%%%%%
The construction of the deformed algebra in~\cite{Kow.Smo:04} 
starts by attempting to put
together a non-commutative
spacetime with a similarly non-commutative energy-momentum
space. The commutation relations initially postulated are
\bae
\left[ M_{\mu \nu},M_{\rho \sigma} \right]
\fe
g_{\mu \sigma} M_{\nu \rho} 
+g_{\nu \rho} M_{\mu \sigma}
-g_{\mu \rho} M_{\nu \sigma}
-g_{\nu \sigma} M_{\mu \rho}
\label{MMcr}
\ff
\left[ M_{\mu \nu},X_{\rho} \right]
\fe
-g_{\mu \rho} X_{\nu}
+g_{\nu \rho} X_{\mu}
\label{MXcr}
\\
\left[ M_{\mu \nu},P_{\rho} \right]
\fe
-g_{\mu \rho} P_{\nu}
+g_{\nu \rho} P_{\mu}
\label{MPcr}
\\
\left[ X_{\mu},X_{\nu} \right]
\fe
\frac{1}{\kappa^2} M_{\mu \nu}
\label{XXcr}
\\
\left[ P_{\mu},P_{\nu} \right]
\fe
\frac{1}{R^2} M_{\mu \nu}
\label{PPcr}
\, ,
\eae
and the question is asked whether the Jacobi identities can be
satisfied. The conclusion reached is that
the canonical commutation relations,
\ble{XPcr}
[X_{\mu},P_{\nu}]= g_{\mu \nu}
\, ,
\ee
have to be deformed. We read:\\[2mm]

\begin{minipage}{.8 \textwidth}
{\em One finds by explicit computation that the Jacobi
identities are satisfied if one takes instead}
\end{minipage}
\ble{xpccrs}
[X_{\mu},P_{\nu}]
=
g_{\mu \nu}
-\frac{1}{\kappa^2} P_{\mu} P_{\nu}
-\frac{1}{R^2} X_{\mu} X_{\nu}
+\frac{1}{\kappa R}
( X_{\mu} P_{\nu} +P_{\mu} X_{\nu} +M_{\mu \nu} )
\, .
\ee
The problem with this statement originates in the
form Eq.~(\ref{XPcr}) is written. Since we are dealing with a Lie
algebra, the right hand side of that equation ought to be
linear in the generators of the algebra. The equation is
nevertheless generally written in that form because the
corresponding generator, call it $F$, is central in the
algebra. Writing the Jacobi identities with only a $g_{\mu
\nu}$ in the \rhs{} of (\ref{XPcr}) amounts to assuming that
$F$ remains central after the deformation, which
is not true. Indeed, taking $F$ into account, one finds from
the Jacobi identity for the nested commutators
$[X_{\rho},[X_{\mu},P_{\nu}]]$, $[P_{\rho},[X_{\mu},P_{\nu}]]$
that
\ble{FXPcr}
[X_{\rho},F]=\frac{1}{\kappa^2} P_{\rho}
\, ,
\qquad
\qquad
[P_{\rho},F]=-\frac{1}{R^2} X_{\rho}
\, ,
\ee
where
\ble{XPFcr}
[X_{\mu},P_{\nu}]=g_{\mu \nu} F
\, ,
\ee
and $F$ still commutes with the Lorentz sector. 
The resulting Lie algebra, given by
Eqs.~(\ref{MXcr})--(\ref{PPcr}), (\ref{FXPcr})
and~(\ref{XPFcr}) (all other commutators being zero) is a
3-SSR Lie algebra, with invariant velocity, mass and length scales
set by $c=1$, $\kappa$ and $R$, respectively --- we denote it
by $\calG_{\kappa, R}$ in what follows%
\footnote{%
The standard practice seems to be to exclude $\hbar$ from the
counting of invariant scales --- we follow it here to avoid
confusion.%
}. 

Despite the
obvious aesthetic and practical disadvantages,
one can, in principle, choose to close the algebra
non-linearly, maintaining $F$ central, as in~\cite{Kow.Smo:04}.
Given that the main motivation for doing so in the first place was
the introduction of an additional invariant scale, and that
this can be achieved, as shown above, without sacrificing linearity, 
there seems to be hardly any reason for making that choice. 
Additionally, one would then have to also supply a coalgebra 
structure, \ie, a
rule about how does the algebra act on tensor products of
representations, satisfying reasonable {\em physical}
requirements --- without it, no composite physical systems
can be considered and the usefulness of the deformation is
drastically reduced. On the other hand, maintaining the 
deformation within the
Lie algebra framework, has the added advantage that the
standard coalgebra structure is available, namely, all
generators act on tensor products as derivations.  
%%%%%%%%%%%%%%%%%%%%%%%%%%%%%%%%%%%%%%%%%%%%%%%%%%%%%%%%%%%%%%%%
%%%%%%%%%%%%%%%%%%%%%%%%%%%%%%%%%%%%%%%%%%%%%%%%%%%%%%%%%%%%%%%%%
\section{The Stability Point of View}
\label{TSPOV}
%%%%%%%%%%%%%%%%%%%%%%%%%%%%%%%%%%%%%%%%%%%%%%%%%%%%%%%%%%%%%%%%
%%%%%%%%%%%%%%%%%%%%%%%%%%%%%%%%%%%%%%%%%%%%%%%%%%%%%%%%%%%%%%%%%
{\em Stable} Lie algebras are isomorphic to all Lie algebras with 
infinitesimally
differing structure constants, {\em unstable} ones are not. 
When the structure constants of a 
particular Lie
algebra used in physics involve experimentally determined
quantities, \eg{}, fundamental constants, then it is natural
to seek a stable form of the algebra, in order to guarantee the
robustness of the associated physics. This point of view has
already a long history (see, for example,
\cite{Bay.Fla.Fro.Lic.Ste:78,Fla:82,Vil:94}) and can
boast at least two (alas, {\em a posteriori}) predictions: (i) the
algebra of Galilean kinematics is unstable --- its stabilized
form is relativistic kinematics (with the particular value of
$c$ fixed, of course, by experiment) and (ii) the infinite
dimensional algebra of functions on classical phase space,
with Lie product given by the Poisson bracket, is unstable,
its stabilized form being quantum mechanics (with $\hbar$
fixed by experiment). Faddeev has pointed out~\cite{Fad:88} 
that a similar
relation exists between special and general relativity, the
gravitational constant $G$ being the deformation parameter in
this case. 

The mathematical aspects of the stability problem have been
worked out in the classical contributions of
Gerstenhaber~\cite{Ger:64},
Nijenhuis and Richardson~\cite{Nij.Ric:66,Nij.Ric:67}, and others.
There exists a beautiful cohomological description of the
tangent space $T\calL_n$ to the space $\calL_n$ of all Lie 
algebras of a certain
dimension, with directions leading to non-isomorphic Lie
algebras being associated to non-trivial cocycles of a certain
coboundary operator. This formulation, coupled with Whitehead's
lemma, shows that semisimple Lie algebras are stable.
In~\cite{Vil:94}, the formalism was applied to the problem of
determining the stable form of the Poincar\'e algebra,
extended by the inclusion of the coordinates as generators.
The result is, up to trivial rescalings and the additional 
freedom of certain signs, the 3-SSR Lie algebra
$\calG_{\kappa,R}$ of 
the previous section%
\footnote{%
A second stable algebra found in~\cite{Vil:94} seems less
relevant physically --- only the $P$-$X$, $P$-$F$, $X$-$F$
relations are deformed (\ie, coordinates and momenta are still
commutative), and the two invariant scales $\kappa$ and $R$ appear 
only as $1/\kappa R$, so that the whole deformation disappears when $R
\rightarrow \infty$.%
}%
. Being
semisimple, it is stable and hence no further
non-trivial deformations are possible, and no new invariant
scales can be introduced. As argued in~\cite{Vil:94}, one may
take the limit $R \rightarrow \infty$ if one is interested in
the kinematics in the tangent space, rather than the motions
in the manifold itself. The resulting algebra
$\calG_{\kappa,\infty}$ is then
unstable, but this is just due to the fact that a flat
manifold (the tangent space) is qualitatively different than a
curved one. Viewed in another way, the resulting algebra is
stable if one restricts to the subspace of $\calL_n$ of tangent space
kinematical algebras.     
%%%%%%%%%%%%%%%%%%%%%%%%%%%%%%%%%%%%%%%%%%%%%%%%%%%%%%%%%%%%%%%%
%%%%%%%%%%%%%%%%%%%%%%%%%%%%%%%%%%%%%%%%%%%%%%%%%%%%%%%%%%%%%%%%%
\section{Relation with Other Algebras}
\label{RwOA}
%%%%%%%%%%%%%%%%%%%%%%%%%%%%%%%%%%%%%%%%%%%%%%%%%%%%%%%%%%%%%%%%
%%%%%%%%%%%%%%%%%%%%%%%%%%%%%%%%%%%%%%%%%%%%%%%%%%%%%%%%%%%%%%%%%
A number of non-commutative spacetime Lie algebras have been
proposed over the years, but rarely did a full set of
commutators (including momenta and the Lorentz sector) appear in 
these proposals. Additionally, non-linear deformations have
appeared, so the task of comparing $\calG_{\kappa,R}$
with earlier approaches is not
straight-forward. We undertake it in detail in our forthcoming
article mentioned above, restricting ourselves to some brief
remarks in this section --- we emphasize that the list of
references that
follows is not exhaustive. 

A non-commutative spacetime algebra, in which the coordinates
generate rotations in a 5-dimensional space of constant
negative curvature, was proposed almost sixty years ago by
Snyder~\cite{Sny:47}. The $x$-$x$ relations there are
identical to the ones in~(\ref{XXcr}). The momenta commute
(which corresponds to the $R \rightarrow \infty$ limit 
mentioned above)
but the $P$-$X$ relations contain non-linear terms. Soon
thereafter, an article by Yang~\cite{Yan:47} pointed out that
by recognizing the additional generator in the \rhs{} of the
$P$-$X$ relations, as we propose in Sect.{}~\ref{TLFTSR}, 
one may render the algebra linear. His momenta cease to
commute, in exactly the same way as in~(\ref{PPcr}), and the
algebra is, up to rescalings, the stable form
$\calG_{\kappa,R}$ of~\cite{Vil:94}.
We choose nevertheless~\cite{Vil:94} as our
main reference, because there, as explained already, the physically 
sensible 
criterion of stability is proposed which, applied to the problem
at hand, uniquely specifies its solution. The particular
commutators in Yang's work were chosen so that the resulting
algebra were $O(1,5)$, without any further justification or
hint of uniqueness.
A later work by Khruschev and
Leznov~\cite{Khr.Lez:02} claimed that Yang's algebra could be
further deformed, and introduced an additional invariant scale
with dimensions of action. It can be easily shown though that
the extra terms there can be reabsorbed by a linear
redefinition mixing coordinates with momenta. 

Apart from the above proposals, non-linear deformations have been
introduced, as mentioned in the 
introduction. Since these involve a deformation of the
coalgebra sector as well, a proper analysis and evaluation
cannot be given here. It is nevertheless interesting to point
out that several of the forms that have been proposed have
been shown to be related by non-linear redefinitions of the
generators~\cite{Kow.Now:02}. The Lorentz-plus-coordinates
sector can be brought in a linear form that coincides with the
same sector of $\calG_{\kappa,R}$, while the momenta commute. In other 
words, apart from the $P$-$X$ relations, the standard 2-SSR
non-linear algebras (termed DSR1 and DSR2 in~\cite{Kow.Now:02}) are 
isomorphic to $\calG_{\kappa,\infty}$, while the
linearized version of the 3-SSR of~\cite{Kow.Smo:04} is, as
shown earlier, $\calG_{\kappa,R}$. That the $P$-$X$ relations
are different in DSR$n$ than those proposed here is hardly
surprising given that in those works too, the extra generator
$F$ passes unnoticed. In any case, in those works the $P$-$X$
relations are derived from a Heisenberg double construction,
in which the coalgebra sector plays a role --- accordingly, we
defer further comments to future work. We think that these brief
remarks suffice to show that stability considerations are
relevant to the problem of multiple-scale relativities and
strongly suggest seeking a solution (or rather, using the one
found long ago) within the standard Lie
algebra framework. 
%%%%%%%%%%%%%%%%%%%%%%%%%%%%%%%%%%%%%%%%%%%%%%%%%%%%%%%%%%%%%%%%
%%%%%%%%%%%%%%%%%%%%%%%%%%%%%%%%%%%%%%%%%%%%%%%%%%%%%%%%%%%%%%%%%
%\bibliographystyle{plain}
%\bibliography{strings}

\begin{thebibliography}{10}

\bibitem{Ahl:03}
D.{}~V.{} Ahluwalia-Khalilova.
\newblock Operational {I}ndistinguishability of {D}oubly {S}pecial
  {R}elativities from {S}pecial {R}elativity.
\newblock 2003.
\newblock \texttt{gr-qc/0212128}.

\bibitem{Bay.Fla.Fro.Lic.Ste:78}
F.~Bayen, M.~Flato, C.~Fronsdal, A.~Lichnerowicz, and D.~Sternheimer.
\newblock Deformation {T}heory and {Q}uantization.
\newblock {\em Ann. Phys.}, 111:61--151, 1978.

\bibitem{Fad:88}
L.~D. Faddeev.
\newblock {\em Asia-Pacific Physics News}, 3:21, 1988.
\newblock Also in ``Frontiers in Physics, High Technology 
and Mathematics''
  (ed. Cerdeira and Lundqvist) p. 238-246, World Scientific, (1989).

\bibitem{Fla:82}
M.~Flato.
\newblock {\em Chech. J. Phys.}, B32:472, 1982.

\bibitem{Ger:64}
M.~Gerstenhaber.
\newblock On the {D}eformation of {R}ings and {A}lgebras.
\newblock {\em Ann. Math.}, 79:59--103, 1964.

\bibitem{Khr.Lez:02}
V.{}~V.{} Khruschev and A.{}~N.{} Leznov.
\newblock The {R}elativistic {I}nvariant {L}ie {A}lgebra 
for the {K}inematical
  {O}bservables in {Q}uantum {S}pace-{T}ime.
\newblock 2002.
\newblock \texttt{hep-th/0207082}.

\bibitem{Kow.Now:02}
J.{} Kowalski-Glikman and S.{} Nowak.
\newblock Non-commutative {S}pace-{T}ime of {D}oubly 
{S}pecial {R}elativity
  {T}heories.
\newblock 2002.
\newblock \texttt{hep-th/0204245}.

\bibitem{Kow.Smo:04}
J.{} Kowalski-Glikman and L.{} Smolin.
\newblock Triply {S}pecial {R}elativity.
\newblock 2004.
\newblock \texttt{hep-th/0406276}.

\bibitem{Vil:94}
R.~Vilela Mendes.
\newblock Deformations, {S}table {T}heories and 
{F}undamental {C}onstants.
\newblock {\em J. Phys.}, A27:8091--8104, 1994.

\bibitem{Nij.Ric:66}
A.{} Nijenhuis and R.W.{} Richardson.
\newblock Cohomology and {D}eformations in {G}raded {L}ie {A}lgebras.
\newblock {\em Bull. Amer. Math. Soc.}, 72:1--29, 1966.

\bibitem{Nij.Ric:67}
A.{} Nijenhuis and R.W.{} Richardson.
\newblock Deformations of {L}ie {A}lgebra {S}tructures.
\newblock {\em J. Math. and Mech.}, 17:89--105, 1967.

\bibitem{Sny:47}
H.{}~S.{} Snyder.
\newblock Quantized {S}pace-{T}ime.
\newblock {\em Phys.{} Rev.}, 71:38, 1947.

\bibitem{Yan:47}
C.{}~N.{} Yang.
\newblock On {Q}uantized {S}pace-{T}ime.
\newblock {\em Phys.{} Rev.{}}, 72:874, 1947.

\end{thebibliography}
%%%%%%%%%%%%%%%%%%%%%%%%%%%%%%%%%%%%%%%%%%%%%%%%%%%%%%%%%%%%%%%%%

\end{document}